\documentclass[openbib,fleqn,12pt]{article}
\usepackage{graphicx}
\newcommand {\beq} {\begin{equation}}
\newcommand {\bea} {\begin{eqnarray} \nonumber }
\newcommand {\eeq} {\end{equation}}
\newcommand {\eea} {\end{eqnarray}}

\newcommand {\al} {\alpha}

\begin{document}
\bibliographystyle{unsrt}
\begin{titlepage}

{\centering
{\Large \bf  DISTRIBUTION OF RESONANCE WIDTHS IN LOCALIZED TIGHT-BINDING 
MODELS.
   \\}
{\Large \bf }
\vskip 2.truecm
{\large
 Marcello Terraneo (${}^{1}$) and Italo Guarneri
(${}^{2}$)\\}
{\normalsize\em  $({}^{1})$,$({}^{2})$  Dipartimento di Scienze Chimiche,
Fisiche e Matematiche
\\Universit\`a dell'Insubria -- Polo di Como\\
Via Lucini 3, I--22100 Como, Italy\\
 $({}^{1})$, $({}^{2})$ Istituto Nazionale di Fisica della Materia,
Unit\`a di
Milano, via Celoria 16, 20133 Milano, Italy \\
$({}^{2})$  and I.N.F.N., Sezione di Pavia, via Ugo Bassi 6, 27100 Pavia, Italy \\
$({}^{1})$  E-mail: Marcello.Terraneo@fisica.unimi.it \\
{\sc Preliminary Version} {as of \today \\}
}
\vskip 20pt
\vskip 2.5truecm
{\centerline {\sc \large Abstract}}}
{\noindent We numerically analyze the distribution of scattering resonance
widths 
in one- and quasi-one dimensional tight binding models, in the  
localized regime. We detect and discuss an algebraic decay of the 
distribution, 
 similar, though  not identical, to recent theoretical predictions.
}

\end{titlepage}
\setcounter{footnote}{0}

\section{Introduction.}

The decay in time of the survival probability
inside open quantum systems is a nontrivial  issue, both in Mesoscopic 
Physics and in Quantum Chaology. Such decay is determined by the distribution 
of resonance widths, which has been studied
extensively\cite{?}.

Still, the effect of   localization on the statistics of scattering 
resonances is not completely 
understood. In the strongly localized regime, some
arguments \cite{CMS}
predict 
an asymptotic ${t^{-1}}$ law 
for the probability decay and a  $\Gamma^{-1}$ behaviour for the 
distribution of resonance widths $\Gamma$; more recently, an 
analytical theory was developed \cite{FT}, which slightly corrects the 
latter into an average decay $\Gamma^{-1.25}$. 

In this paper we numerically address this 
question by investigating the $\Gamma $ distribution in a class 
of quasi-one dimensional models. Hamiltonians in this class are given 
by Band Random Matrices. As such matrices provide models for 
quantum localization not only in disordered solids but in chaotic 
Hamiltonian systems, too \cite{Chir}, our present results are also relevant 
to the latter class of problems \cite{Mas}.

We use a computational 
scheme based on the Effective Hamiltonian (EH) approach: however, at variance 
with the usual way of implementing EHs, which neglects 
their energy dependence, we perform an exact (within the limits 
of numerical accuracy) computation of the distribution. Our method 
is described in sec. 2.
 In this way 
we indeed detect a large interval of $\Gamma$ in which $P(\Gamma)$ decays 
like $\Gamma^{-\alpha}$ with $\alpha $ in the range $1 - 2$, depending
on the
model, and on the localization ratio.
Our results therefore signal algebraic decay, still somewhat 
different from the predictions of ref.\cite{FT}. Some other differences 
appear, in the large-${\Gamma}$ part of the distribution. 

However, comparing numerical results with theoretical predictions is by no 
means an obvious task. First, the theory of ref.\cite{FT} is one 
for a {\it continuous} model, while our models are discrete. Second, 
that theory somehow assumes a certain ideal coupling to continuum, different 
from ours. Finally,  the large-size asymptotic regime in this problem 
has some nontrivial features, which impose caution   
in analyzing finite-sample data.

In sec.3 we discuss some general features 
of the problem, based 
on general facts about Anderson localization, and on 
elementary mathematical estimates which we derive for our class of discrete 
models, and which are well confirmed by our data. 
Our numerical results are described and discussed in the conclusive sec. 4.

\section{Models, and Effective Hamiltonians.}

We consider  one-dimensional lattice Hamiltonians, which describe a wire 
coupled to one or two perfect leads, in the form:
\begin{equation}
H=H^{(i)}+H^{(o)}+H^{(io)}
\end{equation}
The first two terms describe the wire and the leads,
respectively;  the third term describes the coupling between them.
We label lattice sites by an integer  $j$, in such a way that the wire 
is described by $1\leq j\leq N$. The Hamiltonian in the leads describes 
hopping between sites spaced by a fixed integer $b$; 
in the basis of vectors $|j\rangle$, it has nonzero matrix elements 
only between sites $i,j$ which belong in the same lead, that is, 
they are both larger than $N$  or smaller than $1$. In that case,  
\begin{equation}
\label{ham}
H^{(o)}_{ij}=\delta_{i,j+b}+\delta_{i,j-b}
\end{equation}
The wire Hamiltonian is a finite, $N\times N$ matrix   with nonzero elements 
only between sites within the wire. Finally, the  operator 
$H^{(io)}$ couples sites $i,j$ spaced by $b$, one lying in the wire
and the other in a lead. The corresponding matrix elements again have the 
form (\ref{ham}).     
 
We have considered two special cases, namely:

\noindent
(i) $b=1$, $H^{(i)}$ a tridiagonal matrix with unit off-diagonal elements 
and diagonal elements given by independent random variables uniformly 
distributed in the interval $[-W/2,+W/2]$. This is a finite sample of a 
one-dimensional Anderson model coupled to leads on both sides.
In the discussion below we shall also make reference to the one-sided 
Anderson model, in which the Hamiltonian matrix is a semi-infinite rather 
than a doubly infinite one. This is equivalent to inserting a perfectly 
reflecting boundary at $n=N+1$.
\par\noindent
(ii) $N>b>1$,  the wire Hamiltonian is a Band Random Matrix (BRM), that is,
a
 real symmetric matrix of
 rank $N$ such that $H_{i,j} \ne 0 \leftrightarrow |i-j| \le b$. 
$H_{i,j}^{(i)}$ are independent Gaussian variables, with
 variance $\frac{\beta^2}{2}$ for off-diagonal elements and $\beta^2$ for
diagonal
 ones.
We have chosen $\beta=1$. As analyzed in \cite{CGM}, this corresponds to
the ``matching wire'' regime discussed by Ekonomou and
 Soukolis \cite{ES}.

In both cases (i),(ii) the Schroedinger equation for free propagation in the 
leads 
has solutions $u_m = \frac{1}{\sqrt{2\pi}} e^{ikm}$, with 
dispersion law $ E = 2 \cos kb$.
 For any energy value $E$ in the interval $[-2,2]$ there are $b$ different 
allowed momenta, 
\beq
k_s = \frac{\arccos {\frac{E}{2}}}{b} + \frac{2\pi}{b} s
\eeq
with $s =0 \ldots b-1$. There are $b$ incoming and $b$  outgoing 
 waves, hence the wire enforces multichannel scattering. 
It is important to remark that  with the choice (\ref{ham})
 the velocity $ v_i = \frac{d H}{d k_i} | _{H=E}$ is the same in all
channels\footnote{In ref. \cite{CGM} it was noted that the statistics
of conductance fluctuations is not significantly different with other
choices of the dispersion law.}.

 The S-matrix relates amplitudes of incoming and outgoing plane waves,
 $I_{L,R}$ and $O_{L,R}$ respectively ($L$ and $R$ stand for left and right):
\[
S    \left(
\matrix{I_L \cr I_R} \right) 
   =    \left( \matrix{O_L \cr O_R} \right)
\]
In our representation, the S-matrix is a $2b \times 2b$ energy-dependent 
matrix $ S_{s,l} (E)$, where $l$, $s$ are channel indexes.

We compute the scattering matrix from the Lippman-Schwinger equation for
 the scattering states
\beq
u^\pm-G_0^\pm\cdot V u^\pm = u
\eeq
where $u$ are the free eigenfunctions, $u^\pm$ the scattering states and V
is
 the ``potential'': a matrix of rank N defined via the formula $H - H_0 = V$.
$G_0^\pm $ is the free Green function $G_0^\pm = (E \pm i\epsilon
 - H_0)^{-1}$. It  can be computed  by  a complex  integral: 

\begin{eqnarray} \label{inte} 
                  (G_0^\pm)_{n,m} & =\langle 
                  n|(E-H_0{\pm}i \epsilon)^{-1}|m\rangle 
                  \big|_{\epsilon=0}  
                  & = {1\over {2 \pi}} \int_0^{2 \pi} dk 
                  {e^{i(m-n) k}\over{E{\pm}i \epsilon
                  -{2\cos{kb} }}}
                  \Bigg|_{\epsilon=0}    
 \end{eqnarray}

The scattering matrix is given by:
 \beq 
S_{i j}=\delta_{i j}- 
2 \pi i \sqrt{{dk_i \over dH}} \Bigg|_{H=E} \sqrt{{dk_j \over dH}}\Bigg|_{H=E} 
<  u_i \left| V \right| u_j^\pm >
\eeq
where $i$, $j$ are channel labels, and $\sqrt{{dk_i \over dH}} \Bigg|_{H=E}$
is the density of states in channel $i$.

Now, $G^0$ can  be written in  block form as
$$
\left[
\matrix{
g_{11}^0 & g_{12}^0 & g_{13}^0 \cr
g_{21}^0 & g_{22}^0 & g_{23}^0 \cr
g_{31}^0 & g_{32}^0 & g_{33}^0
}
\right] 
$$
In this section  we use small letters for blocks ({\it i.e}, submatrices) and capital letters 
 for full operators.
Here $g_{11}^0$, $g_{33}^0$, $g_{31}^0$ and $g_{13}^0$
are semi-infinite
matrices, $g_{22}^0$ is a $N \times N $ matrix, $g_{12}^0$ and $g_{32}^0$
are matrices with $N$ columns and infinitely many rows, 
$g_{21}^0$ and $g_{23}^0$ have $N$ rows and infinitely many columns.
As the potential matrix is localized within 
the center block (2,2), we write it 
in the block form
$$
\left[
\matrix{
0 & 0 & 0 \cr
0 & v & 0 \cr
0 & 0 & 0
}
\right] 
$$

The poles of the S-matrix are the complex values of energy for which the 
$N\times N $ matrix $(I -
g_{22}^{0\pm}
v) $ has no inverse; hence, they are given by the roots of
\beq
det(I-g_{22}^{0\pm}v) = 0 
\label{det}
\eeq 
In order to solve this  equation we introduce an 'effective Hamiltonian' 
 matrix $H_{eff}$ of rank N. This is a well known construction
 \cite{Verba}, \cite{Guhr},
 \cite{Beena},
but the exact form of the effective hamiltonian 
for  our specific models is 
not immediately derived  from the general theory.   
 There are in fact certain  slight differencies
between our effective Hamiltonian and  the one used in   
\cite{FT}, which are probably due to a different choice of the Hamiltonian in
the leads. Therefore we shall presently give a complete derivation for our 
specific models. 

We start with  the identity
\begin{equation} \label{ls}
G_0 = G - G_0 V G = (1 - G_0 V ) G.
\end{equation}
where $G = (E\pm i\epsilon - H)^{-1}|_{\epsilon = 0} $
Multiplying on the left by $G_0^{-1}$ and on the right by $G^{-1}$, we have
$$ G^{-1} = G_0^{-1} -V .$$

 The  block form of (\ref{ls}) is
$$
\left[
\matrix{
g^0_{11} & g^0_{12} & g^0_{13} \cr
g^0_{21} & g^0_{22} & g^0_{23} \cr
g^0_{31} & g^0_{32} & g^0_{33}
}
\right]  
 =
\left[
\matrix{
1 & - g^0_{12} v & 0 \cr
0 & ( 1-g^0_{22} v ) & 0 \cr
0 & - g^0_{32} v & 1
}
\right]
\left[
\matrix{
g_{11} & g_{12} & g_{13} \cr
g_{21} & g_{22} & g_{23} \cr
g_{31} & g_{32} & g_{33}
}
\right] 
$$
where $1$ and $0$ are identity  and zero matrices of 
dimensions
corresponding to their position in the infinite matrix.

For the center block  we have
$ g_{22}^0 = ( 1 - g_{22}^0 v ) g_{22}$, whence, 
 multiplying on the left by ${g_{22}^0}^{-1}$ and on the right
by $g_{22}^{-1}$, we obtain 
\begin{equation} \label{ls2} 
g_{22}^{-1} = ({g_{22}^0}^{-1} - v ) .
\end{equation}
In order to  compute $(g_{22}^0)^{-1}$ we start from the definition,  
which in block forms reads 
$$
\left[
\matrix{
g^0_{11} & g^0_{12} & g^0_{13} \cr
g^0_{21} & g^0_{22} & g^0_{23} \cr
g^0_{31} & g^0_{32} & g^0_{33}
}
\right]  
\left[
\matrix{
E - h^0_{11} & u_{12} & 0 \cr
u_{21} & E-h^0_{22} & u_{23} \cr
0 & u_{32} & E-h^0_{33}
}
\right]  
=
\left[
\matrix{
1 & 0 & 0 \cr
0 & 1 & 0 \cr
0 & 0 & 1
}
\right]
$$
where $u_{ij}=-h^0_{ij}$ are the coupling matrices.

The equation for the center block is 
\begin{equation} \label{keyeq}
   g^0_{21} u_{12}+ g^0_{22} ( E- h^0_{22})+ g^0_{23} u_{32} = 1 .
\end{equation}

From (\ref{inte}) we know that the
free Green
function is a Toeplitz matrix,
whose center block  $g^0_{22}$ can be written:
$$
\left[
\matrix{
x_0 & x_1 & \ldots & x_{N-1} \cr
x_1 & \ddots & \ddots & \vdots \cr
\vdots & \ddots & \ddots & x_1 \cr
x_{N-1} & \ldots & x_1 & x_0
}
\right]
$$ 
where  $ x_{s}=0 $ if $s$ is not a multiple of $b$, while, if $s=lb$, then 
 $ x_{lb}= x_{0} e^{ilk}$, with  $ k= \arccos{(\frac{E}{2})}$.
Keeping this in mind we get
$$( g^0_{21} u_{12} + g^0_{23} u_{32} ) =  
\left[
\matrix{
x_b    & \ldots & x_1    & 0       & \ldots & 0 & x_{N+1} & \ldots &
x_{N+b} \cr
\vdots &        & \vdots & \vdots  &        & \vdots & \vdots & & \vdots \cr
x_{N+b} & \ldots & x_{N+1} & 0     & \ldots & 0 & x_1 & \ldots & x_b
}
\right]
= -  g^0_{22}  \cdot K_b .
$$ 
where $K_b= diag(e^{i k b}, \ldots , e^{i k b},
0, \ldots , 0 ,e^{i k b}, \ldots , e^{i k b} )$ is the ``self-energy''.

Then from (\ref{keyeq}) we obtain: 
$$
(g^0_{22} )^{-1} = ( E - h^0_{22} -K_b ).
$$ 
Finally, (\ref{ls2}) yields: 

$$ (g_{22})^{-1} = ( g^0_{22} )^{-1} - v = ( E - h^0_{22} - K_b - v ) =
( E - h_{22} - K_b(E) ) =
 (E - H_{eff} (E)).
$$
where $H_{eff}(E) = h_{22}  +
K_b(E)$ is the effective Hamiltonian matrix of rank $N$. The role of the effective 
Hamiltonian 
emerges on noting that:
$$
I - g^{0}_{22}  v = g^0_{22} ((g^0_{22})^{-1} - v) = g^0_{22} (E - H_{eff}
(E))
$$
which shows that solving eqn.(\ref{det}) is the same as solving the equation:
\begin{equation}
\label{deteff}
det (H_{eff}(E) -E) = 0
\end{equation}
For further use we rewrite $H_{eff}$ in operator notation. In place of 
$h_{22}$ we rewrite $H^{(i)}$: the Hamiltonian 
operator of the sample, with Dirichlet conditions at $n=0$ and $n=N+1$.
Then 
\begin{equation}
\label{opnot}
H_{eff}(E)=H^{(i)}+
e^{ik(E)b}\left\{\sum\limits_{n=1}^{b}
+\sum\limits_{n=N-b+1}^{N}\right\}|n\rangle\langle n|
\end{equation}
 The leads only affect diagonal
 elements, at sites near the contact points. For the Anderson model,
 where hopping only occurs between neighbouring sites, only the first
 and the last diagonal elements of the Hamiltonian are affected. 
Some  straightforward 
modifications to the above construction are necessary in the 
one-sided Anderson case; we omit details here.
 
Solving the nonlinear eqn.(\ref{deteff}) is a difficult task. When using the 
effective hamiltonian formalism, one typically neglects the dependence 
of $H_{eff}$ on energy, so the problem is reduced to finding eigenvalues 
of the effective hamiltonian at a chosen fixed value $E_0$ of the energy. 
Such eigenvalues depend on $E_0$ as a parameter; for convenience of 
language, we will term them {\it parametric resonances} in the following, 
reserving the name {\it exact resonances} to solutions of eqn.(\ref{deteff}).

\section{ Theoretical Premises.}

A few remarks are in order, about the mathematical problem set by 
the above formalism.  These are most simply formulated for the 
one-sided Anderson case, so we restrict to that case; 
nevertheless, similar arguments can be developed 
for the two-sided Anderson and for the band matrix models. 

If $E$ is a root of eqn.(\ref{deteff}), then there is a 
vector $|\psi\rangle$ of unit norm, satisfying the equation:
\begin{equation}
\label{eigeq}
H^{(i)}|\psi\rangle+e^{ik(E)}|1\rangle\langle 1|\psi\rangle=E|\psi\rangle
\end{equation}
where $H^{(i)}$ is now the Anderson Hamiltonian with Dirichlet boundary
conditions 
at $n=0$, $n=N+1$. Multiplying eqn.(\ref{eigeq}) on the left by $\psi$,
and 
taking 
imaginary parts,  we get:
\begin{equation}
\label{eigeq1}
{\cal I}m(e^{ik(E)})|\langle 1|\psi\rangle|^2=\Gamma
\end{equation}
where we have set $E=x+i\Gamma$ with real $x,\Gamma$. 

From the dispersion law  the following 
analytical expression of $e^{ik(E)}$ follows:  
\beq
\label{disp}
e^{ik(E)  }  = \frac{E + \sqrt{E^2 - 4}}{2}
\eeq
 In the physical sheet, the square root in eqn.(\ref{disp}) has to be 
chosen such that its imaginary part is opposite in sign to $\Gamma=
{\cal I}m(E)$; then an easy computation shows that 
${\cal I}m(e^{ik(E)})$ is opposite in sign to $\Gamma$, so eqn.(\ref{eigeq1})
cannot have solutions with $\Gamma\neq 0$. In order to solve it,   
we have to continue  $H_{eff}$ into the nonphysical sheet, across the branch 
cut $[-2,+2]$.

 We now address the problem of finding estimates for the largest 
$\Gamma$. We consider  parametric resonances first:   
if $E_0$ is chosen in $(-2,2)$, then eqn.(\ref{eigeq1}) 
immediately sets a sharp, realization-independent  bound on parametric 
$\Gamma$:
\begin{equation}
\label{ineq1}
|\Gamma|\leq |\sin (k(E_0))|=\frac{\sqrt{4-E_0^2}}{2} \leq 1 
\end{equation}  
For exact resonances we can only establish a milder, realization-dependent 
bound. 
Multiplying eqn.(\ref{eigeq}) on the left by $\langle 1|$,  
using eqn.(\ref{eigeq}), and the explicit form of $H^{(i)}$, we get:
\begin{equation}
\label{eigeq2}
\langle 2|\psi\rangle=\langle 1|\psi\rangle\left
(E-e^{ik(E)}-V(1)\right)  
\end{equation}
where $V(1)$ is the random on-site potential at site $1$. 
Since $|\langle  2|\psi\rangle|^2+|\langle 1|\psi\rangle|^2
\leq 1$, using eqn.(\ref{eigeq}) we get the inequality:
\begin{equation}
\label{ineq}
\frac{\Gamma}{{\cal I}m(e^{ik(E)})}\leq \frac{1}
{1+\vert E-e^{ik(E)}-V(1)\vert^2}
\end{equation}
which, given a realization of the random potential, sets an upper bound 
to ${\Gamma}$. At the center of the spectrum and small $V(1)$ this bound  
has  approximately the form $|\Gamma|\leq |V(1)|^{-1}$. In turn, this implies 
that those realizations of the random potential which yield 
$\Gamma$'s larger than a given $\gamma$ have a probability not larger 
than $\sim 1/\gamma$. This bound on the large-$\Gamma$ behaviour 
of the $P(\Gamma)$ distribution of exact resonances is much milder than the 
bound (\ref{ineq1}) for parametric resonances. 
We anticipate that this difference 
is manifest in numerical data (see fig.4, 5 and 6)

In the limit $N=\infty$, both $H^{(i)}$ and $H_{eff}(E_0)$ become 
operators in $\infty-$ dimensional Hilbert space. The latter operator is 
obtained by adding a 
rank-one perturbation to the former, which in turn 
 has a dense point spectrum in the interval $I=[-W/2-2,+W/2+2]$ 
(with 
probability 1). From general 
operator theory it follows,  that in the limit 
$N=\infty$ 
 the  distribution of parametric $\Gamma$'s collapses into  
 a Dirac delta at zero - unlike other scattering statistics, 
(e.g.,  the phase shift distribution), which have a smooth 
limit distribution.  This physically intuitive result should  be valid 
for the distribution of exact resonances, too.

Thus, on increasing $N$, we should expect  the 
leftmost part of the finite-$N$ distribution $P(\Gamma)$ 
to rise, and the rest to gradually subside.
To further illustrate this point, we reformulate 
eqn.(\ref{eigeq}) by projecting it onto the basis of eigenvectors 
$u_n$ of $H^{(i)}$. Denoting $E_n$ the corresponding eigenvalues and 
$\psi_n$ the amplitudes of $\psi$, we get:
$$
\psi_n=-\frac{e^{ik(E)}\langle u_n |1\rangle\langle 1 |\psi\rangle} 
{E_n-E}
$$
whence:
$$
\langle 1 |\psi\rangle=-\langle 1 |\psi\rangle e^{i k(E)}\frac{|\langle u_n |1\rangle|^2}{E_n-E}
$$
As $\langle 1 |\psi\rangle=0$ is excluded, the resonant values of $E$ 
must solve the key  equation:
\begin{equation}
\label{main}
\sum\limits_{n}\frac{p_n}{E_n-E}=-e^{-ik(E)}
\end{equation}
where $p_n=\vert\langle 1\vert u_n\rangle\vert^2$.

At large $N$, the eigenfunctions $u_n$ are exponentially localized, 
with localization lenghts $\xi(E_n)$.
For resonances $E=x+i\Gamma$ with $|\Gamma| \ll 1/N$, a
single-pole 
approximation to the lhs of eqn.(\ref{main}) should be valid, because 
eigenfunctions with $E_n$'s much closer than the average level spacing 
typically have an exponentially 
small overlap; so the sum in (\ref{main}) is dominated by 
a single term\cite{remark}.  
 Hence, the narrowest resonances  
can be assumed to solve
$$
p_n e^{ik(E)}\approx (E-E_n)
$$
with $p_n$ small, so
\begin{equation}
\label{appr}
E\approx E_n+p_ne^{ik(E_n)},~~\mbox{\rm and}~~\Gamma\approx p_n\sin(k(E_n)).
\end{equation}
If we further restrict near the center of the spectrum, then the smallest 
$\Gamma$  come from states $u_n$ localized around sites $n_0(n)$ lying 
in the rightmost  part of 
the sample. For these, $\log (p_n)\approx -2n_0(n)/\xi +\eta_n$, with 
$\xi$ the localization length at the band center and $\eta_n$ a fluctuating quantity of 
order $\sqrt n_0$.

 Thus the distribution of very small $\Gamma$'s  is  ruled by the 
distribution of $p_n$'s. If in addition  
  $\Gamma \ll \exp (-2N/\xi)$, the latter distribution is mainly  
determined by fluctuations of $\eta_n$; assuming a gaussian distribution 
for the latter, one gets that in this region $P(\Gamma)$ has the 
lognormal distribution already well known in this context. Nevertheless, 
this part becomes negligible at large $N$, because it comes 
of a fraction $\sim \xi/N$ of the full set of all resonances. 

Away from this 
extreme region, the $\Gamma$ statistics is more and more  affected  by the 
change in $n_0$. Neglecting $\eta_n$ completely, one 
deduces a dependence $\sim 1/\Gamma$, by a simple  argument already 
reproduced in ref.\cite{FT}. The presence of $\eta_n$ just smoothens  
the cusp of $1/\Gamma$ at $\Gamma\approx\exp(-2N/\xi)$, but the 
$1/\Gamma$ law again re-emerges at larger $\Gamma$.

So finite-N, 
normalized  distributions $P(\Gamma)$
have a peak at $\sim\exp(-2N/\xi)$, of height $\sim \exp(2N/\xi)$. It is 
this very peak which eventually builds the limit $\delta-$ distribution. 
In the range $\exp(-2N/\xi)<\Gamma \ll 1/N$, the above rough
argument 
suggests 
a law $\sim 1/\Gamma$; but it must be mentioned that according to 
Titov and Fyodorov 
(\cite{FT}) this inverse law should be restricted to a smaller range.
Some numerical data about this issue will be  given in the next section.

The large $\Gamma$ region is essentially determined by the coupling 
to continuum, so it should be model-dependent. Nevertheless, it 
is reasonable to assume that the number of resonances involved is constant,  
of order $\xi$; if so, again this tail should subside at large $N$, at 
the rate $\sim \xi/N$.

Finally we use eqn.(\ref{main}) to investigate the reliability 
of parametric resonances as approximations of  exact ones. 
The equation for parametric resonances is obtained from eqn.(\ref{main}) by 
replacing 
in the rhs $E$ by a fixed $E_0$ chosen in $(-2,+2)$. 
An obvious requirement for parametric 
resonances to approximate exact ones is that their dependence on $E_0$ 
be mild. Now,
$$
\left\vert \frac{dE}{dE_0}\right\vert=
\left\vert\sum\limits_{n}\frac{|\langle u_n |1\rangle|^2}{(E_n-E)^2}
\right\vert^{-1}\left\vert\frac{dk(E_0)}{dE_0}\right\vert
\geq\frac{\Gamma^2}{\sqrt {4-E_0^2}}
$$
If a given parametric resonance is to vary little on changing  $E_0$  
 on the order of its width $\Gamma$, it is therefore necessary that:
$$
\Gamma^2 \ll {\sqrt{4-E_0^2}}
$$
which shows that the parametric approximation becomes unreliable close to 
the edges $\pm 2$, and in any case at $\Gamma\sim 1$.

\section{Numerical Method, and Results.}
 
As we are only interested in the statistical distribution of the imaginary parts of solutions of (\ref{deteff}), we don't need to compute them exactly. 
 Instead, we use the fact that the number of zeros of an analytic function $f(z)$ inside a 
closed path is equal to the variation of the phase of the function itself along the path divided by $2\pi $. We have therefore considered rectangular 
regions $\{E: |Re(E)|\leq E_0, -\gamma\leq Im(E)\leq 0\}$ in the lower part of the 2nd Riemann sheet. By numerically 
computing the  phase of $det(H_{eff} - E) $ along the boundaries of such regions we obtained $N(\gamma,E_0)$, 
the number of resonances having real parts in $(-E_0,+E_0)$, and widths not 
larger than $\gamma$. Typically $E_0=0.5$ in our computations. 

Repeating the procedure for different realizations of our random 
Hamiltonians, we obtained the histograms of resonance widths shown in Figs. 1 - 6 (there
$P(\Gamma)$ is the probability density for $\Gamma$ values). 
The numerical procedure is quite heavy, so  we were able to process at most 
$400$ realizations of the BRM model. With the
 Anderson model, computation is faster, so we could 
process up to  $1000$ realizations.

Most of the  distributions of resonance widths $P(\Gamma)$  
 computed by the above 
discussed method decay, at very large $\Gamma$, faster than 
power-like, also because of the difficulty of numerically 
building  good statistics in this poorly populated region.

However, in an intermediate region of values of 
$\Gamma$, the observed decay is algebraic,  
proportional to  $\Gamma^{-\alpha}$. The width of this   region 
depends on the localization ratio $r$, which is proportional to 
$b^2/N$ and to $1/(N W^2)$ for the BRM model and for the Anderson model 
respectively. The region of algebraic decay is very broad in 
 strong localized systems, $r \ll 1 $; it
shrinks  as $r$ is increased, and eventually disappears 
  in the
metallic region $r \gg 1$. 

The behaviour of the exponent $\alpha$ is somewhat different in the BRM and in 
the Anderson model. In the former case, 
 $\alpha$ increases as $r$ increases, going from values about
$1.5 - 1.6$ to values about $2 - 2.1$ (see fig.1).
 Data obtained at fixed $r$ and different 
 values of $N$, $b$ ( with $ \frac {N}{b} \ge 5$, though) show 
 that $\alpha$ only depends on the localization
 ratio in the explored parameter range:  see fig. 2, which shows distributions
 with  approximately  the same      localization ratios and different
$N$, $b$.

We have numerically computed distributions of widths for the one and
two-leads 
Anderson model, too, by solving eqn.(\ref{deteff}) as 
explained 
above.  In this case,  $\alpha$ 
remains more or less constant 
around $1.7 - 1.8$: it doesn't seem to depend on the localization
ratio (fig. 3) in the explored parameter range.
The only effect of decreasing $W$ is the predicted shift of the peak of the 
distribution towards larger values $\Gamma$.

Most of our data do not yet pertain to  
the true asymptotic large-$N$ regime. In fact,  the peak 
in the left-hand part of our data is still relatively broad in comparison 
to the right-hand part, so the asymptotically interesting region 
$e^{-2N/{\xi}} \ll \Gamma \ll 1$ is still somewhat restricted. 
Nevertheless  the region of algebraic decay is already clean, and relatively 
stable 
against variation of the localization ratio. We therefore surmise, that 
in more localized regimes the 
local exponent $\al$ would be found 
to smoothly decrease  with $\Gamma$, tending to $1$ as the peak 
 of the distribution is approached from the right. As we shall presently discuss, we get indications in this 
sense from distributions of parametric $\Gamma$'s, whose computation 
could be pushed to significantly more localized situations than 
accessible to exact computations, based on solving 
 eqn.(\ref{deteff}). 

Parametric distributions were  obtained by diagonalizing 
$H_{eff}(0)$. 
Both for the BRM and the Anderson model, they  
exhibit a cut-off at $\Gamma\sim 1$ which is absent in the 
real distributions.  The latter in fact decays to
zero much more gently.
In other words: in all the models we have studied,  
neglecting the energy dependence of effective Hamiltonians 
yields acceptable results only 
at resonance widths appreciably smaller than $1$ (figs. 4, 5 and 6). The
reason 
is that,  
at such large widths, the effective hamiltonian is significantly changing 
already over the width of single resonances. 
These numerical findings are fully consistent with the analysis in 
section 3. 

Exact and parametric distributions fairly well agree in the central part, 
on the right of the peak, but they are again somewhat different on the left. 
The reason is probably that our exact distributions collect 
resonances from a relatively broad interval of real energies, so they 
include resonances closer to the edges, where smaller $\Gamma$'s are 
{\it a priori} expected (eqn.(\ref{appr})).

The sharp  cut-off of numerical distributions of {\it parametric}  
resonances  was also observed 
 in ref.\cite{FT}, and found to be consistent with the therein 
developed theory for {\it exact} resonances in a continuous, 
white-noise model.
That coincidence between exact and parametric 
distributions at relatively large $\Gamma$ raises an interesting 
theoretical problem about the role of the coupling.\cite{FT1}

In fig.4  we compare parametric and exact distributions both for 
the one- and the two-sided Anderson model.  Note that one-lead 
data refer to a significantly more localized regime than 
two-lead ones, and yield smaller $\al$'s. The estimated $\al$ for 
one-sided Anderson is in this case $\approx 1.1$. In fig.5 we have 
again computed the parametric distribution for one-sided Anderson, this time 
in an even more localized situation.  Moving from left to right, 
the lhs part of the distribution 
now exhibits $\al\approx 1$ over three decades, after which 
the distribution gradually 
drops to zero in roughly two decades. 

We finally note that the rightmost part 
of the exact distribution in fig.5 is well fitted by a $\Gamma^{-2}$ law. 
As the latter  coincides with the upper bound established in sec.3 for exact 
resonances, we have an indication  that that bound  is probably optimal.

In summary: in this paper we have analyzed the distributions of imaginary
parts of resonances in different tight-binding  models. The most
interesting result is the presence of a region with  power like
decay, both for Anderson and BRM model. It was obtained by 
numerically  
implementing the effective Hamiltonian method, in a way which doesn't 
{\it a priori} neglect energy dependence. Our analysis indicates that most 
of the distribution of widths tends to concentrate within a single peak 
around $\Gamma\sim\exp(-2N/{\xi})$ 
as more and more localized regimes are approached. While subsiding 
(with increasing $N$), 
the distribution of widths on the right of the peak displays  
an algebraic dependence on $\Gamma$. The related exponent ranges 
from  $1$ (at small $\Gamma$), 
 to $2$ (at large $\Gamma$). In the latter region, however, the specific 
form of the coupling to continuum plays a role.

 Useful discussions with G.Maspero are acknowledged. We are grateful 
to Y.Fyodorov and M.Titov for communicating us unpublished details 
of their work.

\begin{figure}[!h]
\begin{center}
\includegraphics[angle=270, width=8cm]{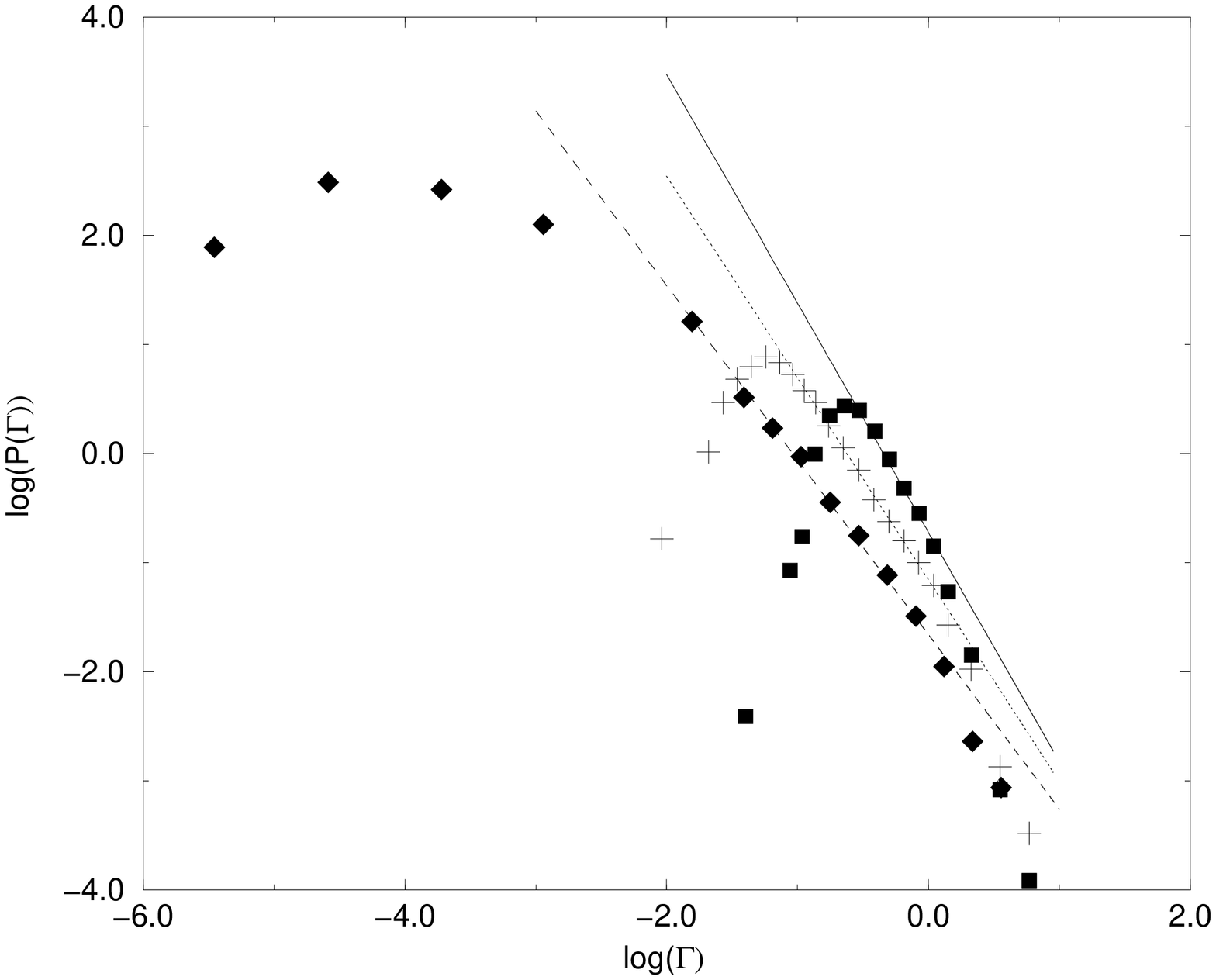}
\end{center}
\caption{P($\Gamma$) for BRM model} \label{a1}
\end{figure}

 \noindent Figure \ref{a1}: {Distribution of exact resonance 
widths  for BRM models 
with
different
localization
 ratios.  Diamonds: $N = 100$, bandwidth $5$, the slope
of the dashed line is $-1.6$.  Squares:  $N=150$, bandwidth $26$, the slope 
of the
 solid  line is $-2.1$. Crosses:  $N = 200$, bandwidth $20$,  slope
$-1.85$.}

\begin{figure}[!h]
\begin{center}
\includegraphics[angle=270, width=8cm]{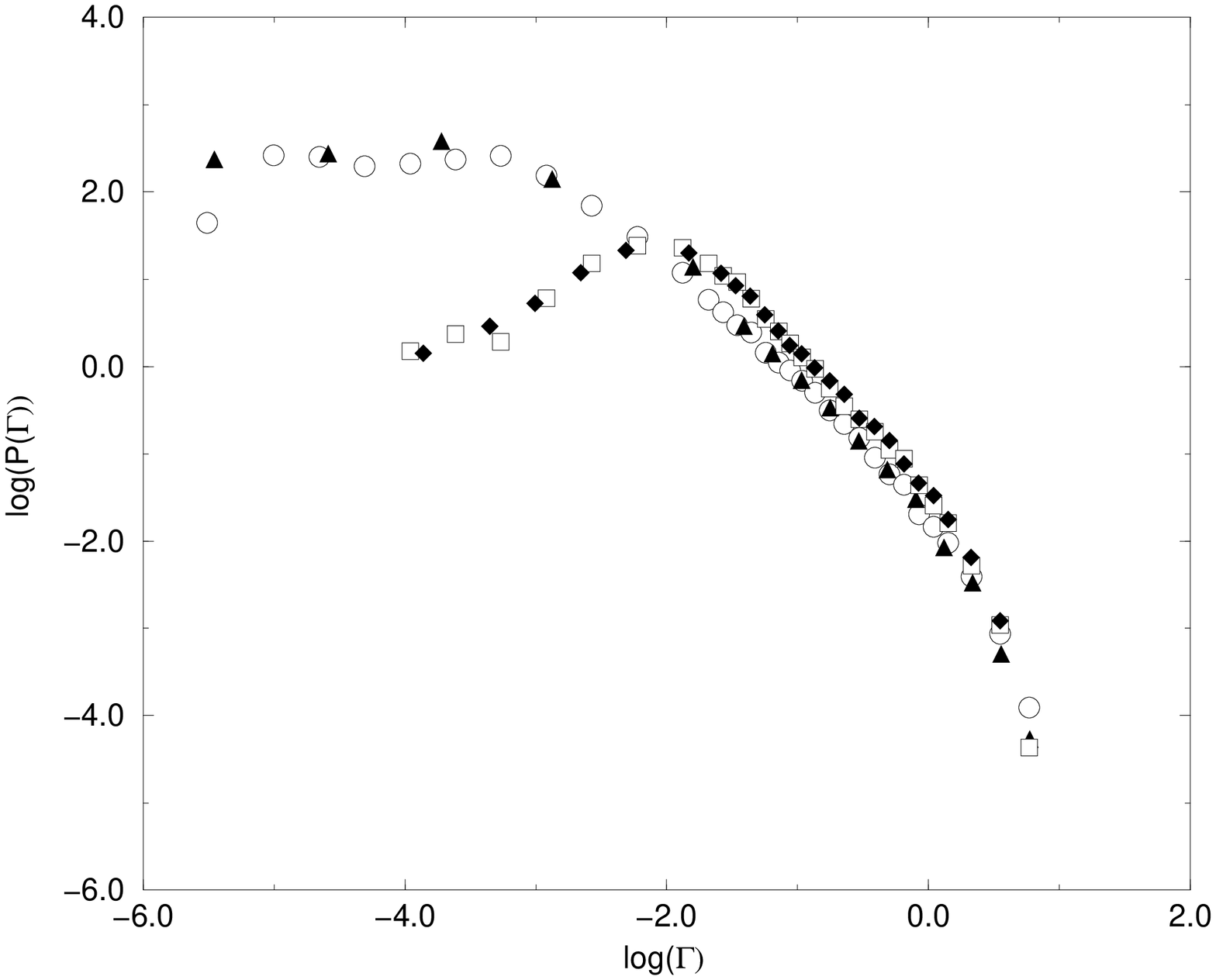}
\end{center}
\caption{P($\Gamma$) for BRM model} \label{a2}
\end{figure}

\noindent Figure \ref{a2}: {Distribution of exact 
resonance widths for BRM models with
similar
localization
 ratios.  Diamonds: $N = 150$,  bandwidth $10$.
Squares: $N=200$, bandwidth $12$. Circles:  $N = 200$,
bandwidth $7$. Triangles: $N=150$, bandwidth $6$. Their  localization ratios
$\frac{b^2}{N}$ are  $0.67, 0.72, 0.245, 0.24$, respectively. }

\begin{figure}[!h]
\begin{center}
\includegraphics[angle=270, width=8cm]{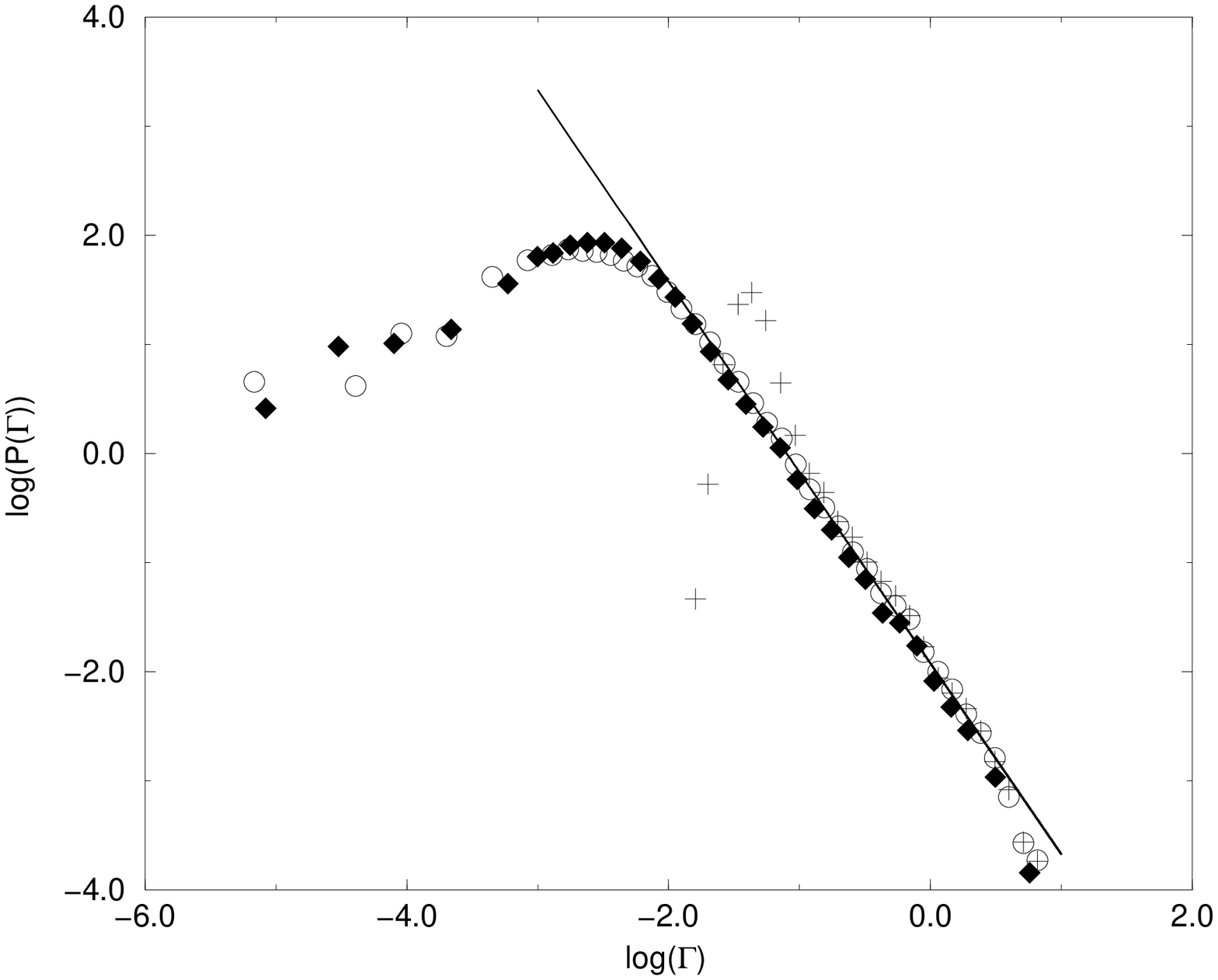}
\end{center}
\caption{P($\Gamma$) for the Anderson model} \label{a3}
\end{figure}

\noindent Figure \ref{a3}: {Distribution of exact resonance widths for the  
two-leads 
Anderson
model,  with
different
 localization ratios.  Diamonds: $N = 150$, $W = 1.1$.
Circles: $N = 100$, $W = 1.5$. Crosses: $N = 100$, $W =
0.3$. The slope of the  line is $-1.75$.}

\begin{figure}[!h]
\begin{center}
\includegraphics[angle=270, width=8cm]{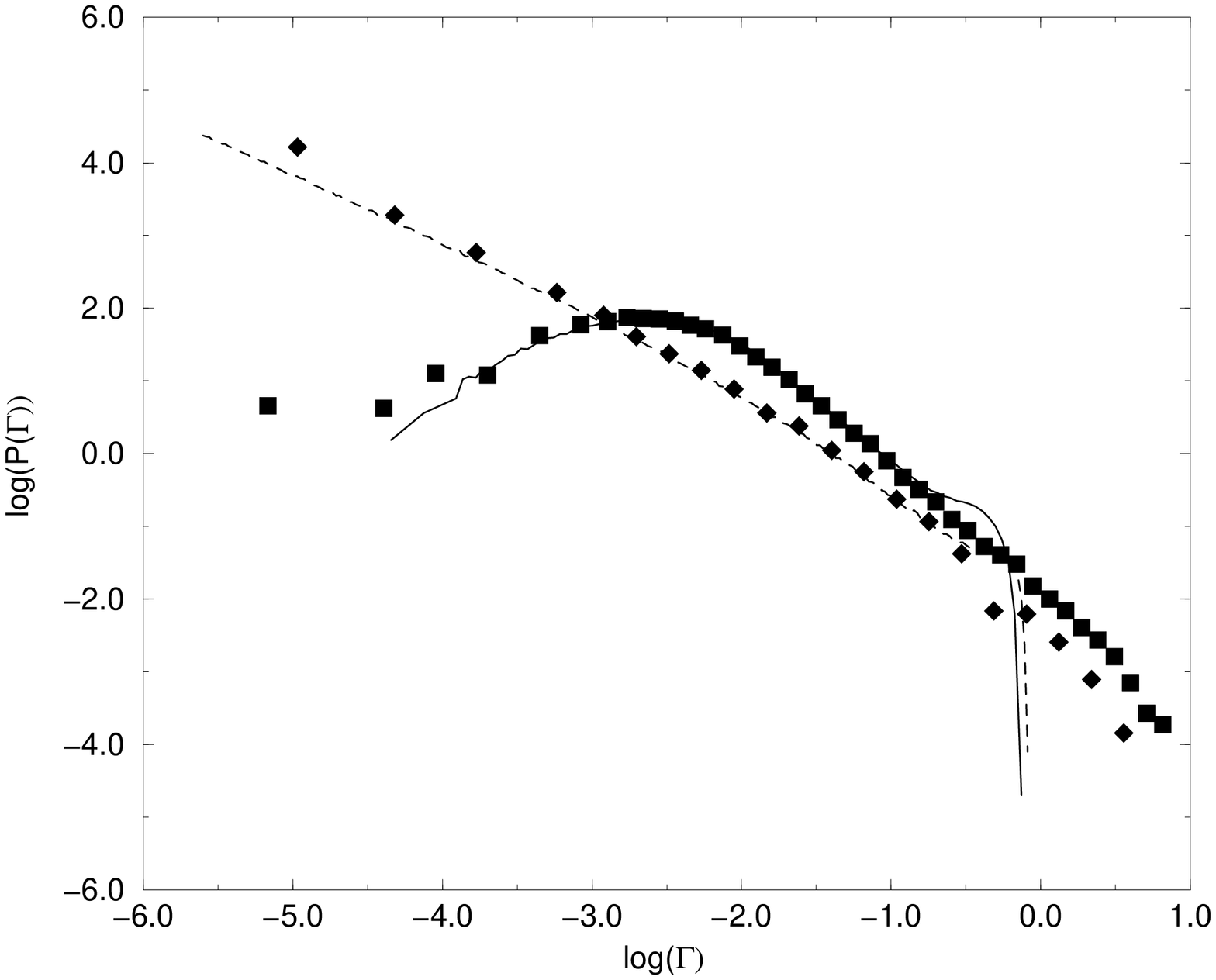}
\end{center}
\caption{ P($\Gamma$) for the Anderson model} \label{a4}
\end{figure}

\noindent Figure \ref{a4}: {
Distribution of parametric and exact resonance widths  for the  Anderson
model. 
Diamonds: exact resonances at $N
= 100$, $W = 2.5$, one lead. Squares: exact resonances at $ N =
100$, $W = 1.5$,  two leads. The dashed and the solid lines 
represent parametric resonances  obtained by 
 diagonalizing  $H_{eff}(0)$,
for 1- and 2- leads model, respectively.}

\begin{figure}[!h]
\begin{center}
\includegraphics[angle=270,width=8cm]{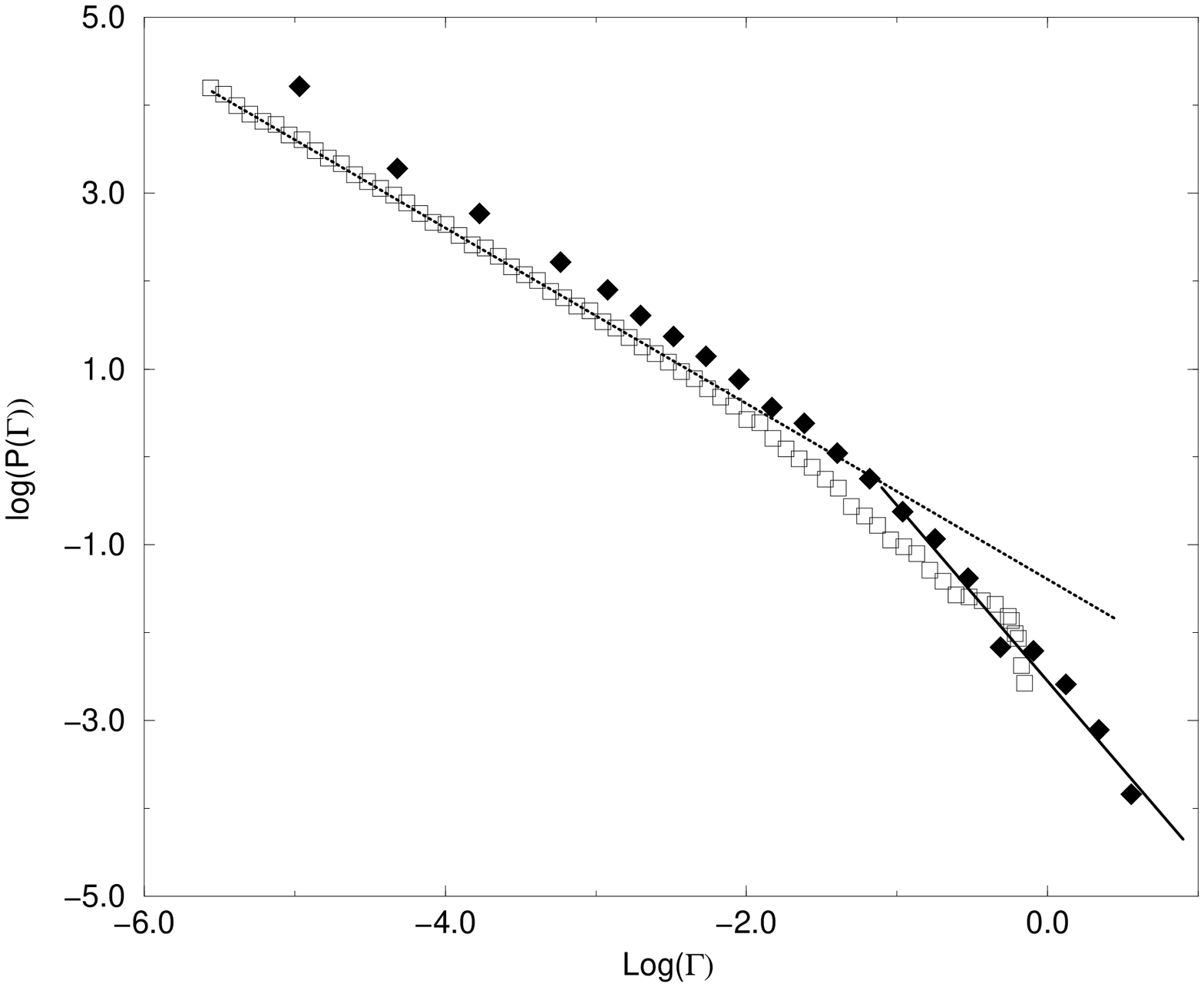}
\end{center}
\caption{P($\Gamma$) for the 1-lead  Anderson model}\label{a6}
\end{figure}

\noindent Figure \ref{a6} {
Distribution of parametric and exact resonance widths  for the 
one-lead Anderson
model. Squares: 
 parametric resonances at 
$N=300$, $W=2$. The slope of the main straight line is $-1$, that 
of the shorter one on the right is $-2$. Diamonds are reproduced 
from Fig.\ref{a4}, which refers to a less localized 
situation. }  

\begin{figure}[!h]
\begin{center}
\includegraphics[angle=270, width=8cm]{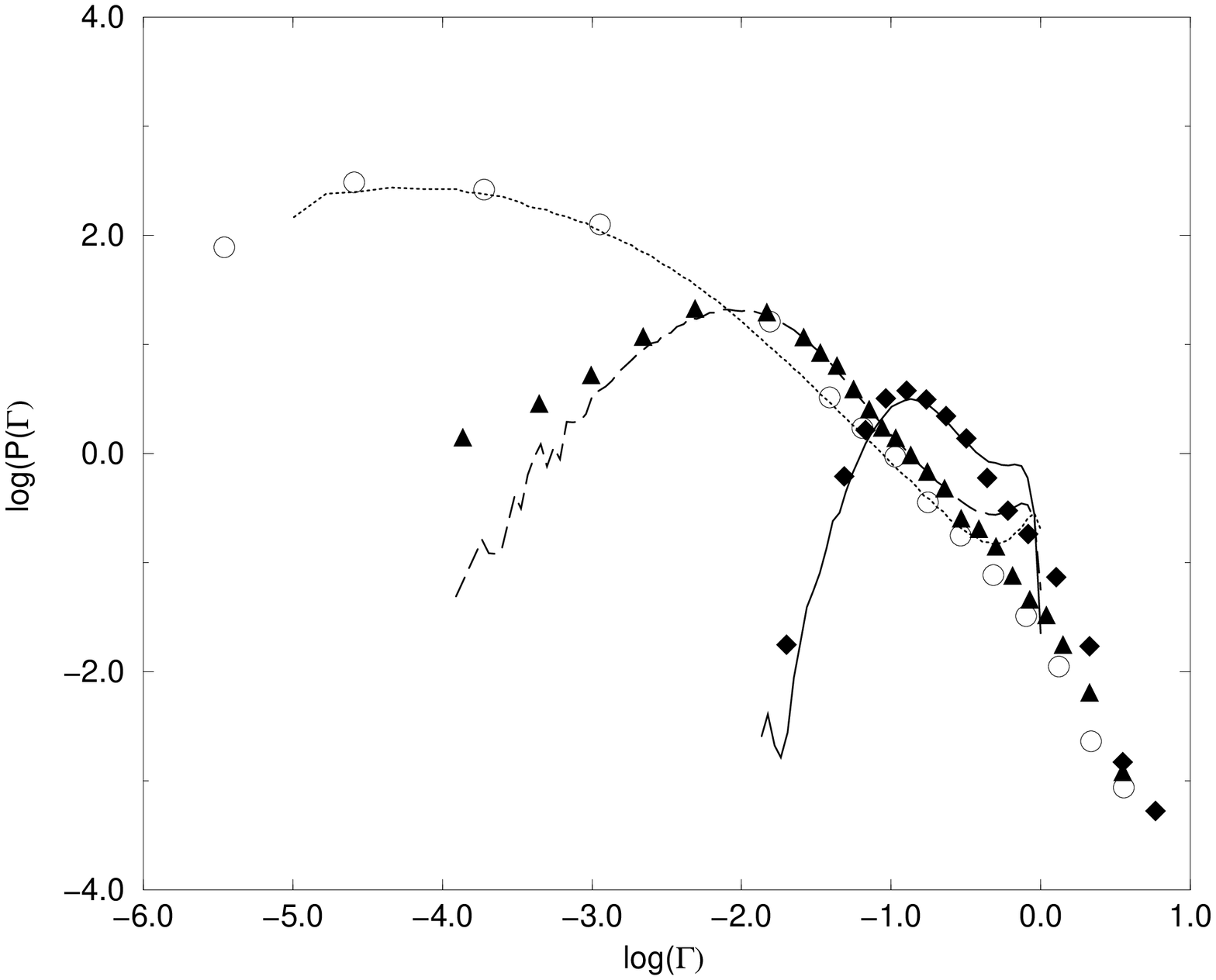}
\end{center}
\caption{P($\Gamma$) for BRM model} \label{a5}
\end{figure}

\noindent Figure \ref{a5}: {Distribution of parametric and exact  resonance 
widths for 
the BRM model.
Diamonds: $N
= 100$, 
bandwidth $15$. Circles: $N = 100$, bandwidth $5$. Triangles, 
$N = 150$, bandwith
$10$. The curves represent parametric resonances obtained by 
diagonalizing $H_{eff}(0)$.} 
\newpage

\newpage

\newpage

\begin{thebibliography}{20}
\bibitem{Mas} Indications in this sense have been found
by G.Maspero, who has detected an algebraic tail $\sim \Gamma^{-1.5} $ with
a dynamical model with localization ratio $\frac{1}{18}$  (private
communication).
\bibitem{remark} We note in passing that the single-pole approximation is 
never valid at $N=\infty$.
\bibitem{FT1} The parametric distributions of ref.(\cite{FT}) 
were obtained by diagonalizing an effective Hamiltonian, which simulates 
coupling to continuum by adding an imaginary shift $i\pi$ to the last 
diagonal element. That marks a difference with our computations, which 
(of necessity) have a $i$ shift instead.  Distributions 
computed that way 
are different from ours in the right hand tails - for instance, they 
exhibit an isolated group of resonances beyond the cutoff, at 
$\Gamma\sim\pi$- but are quite similar at small $\Gamma$.
\end{thebibliography}
\end{document}